\documentclass[12pt]{iopart}
\usepackage{textcomp}
\usepackage[utf8]{inputenc} 
\usepackage{amssymb}
\usepackage{amstext}
\usepackage{graphicx}          
\usepackage{subfigure}
\usepackage{bm}
\usepackage{color}
\usepackage{gensymb}
\usepackage{wasysym}
\usepackage{xcolor,soul}
\usepackage{url}
\usepackage{xr}
\usepackage{pdfpages}
\begin{document}
 
\title{Ferroelectric control of magnetic coupling in Cr dimers adsorbed on $\alpha$-In$_2$Se$_3$}

\author{Jhon F. Ospina$^{1,2}$, S. Di Napoli$^{1,2}$ and  M.A. Barral$^{1,2}$} 

\address{$^1$ Departamento de F\'{\i}sica de la Materia Condensada, GIyA-CNEA, 
Av. Gral. Paz 1499, B1650KNA San Mart\'{\i}n, Prov. Buenos Aires, Argentina}

\address{$^2$ Instituto de Nanociencia y Nanotecnolog\'{\i}a CNEA - CONICET, 
Nodo Constituyentes, Av. Gral. Paz 1499, B1650KNA San Mart\'{\i}n, Prov. Buenos Aires, Argentina}

\ead{andrea.barral@conicet.gov.ar}

\begin{abstract}

Ferroelectric substrates offer a route to controlling magnetic interactions in low-dimensional systems through polarization-dependent changes in local bonding and electronic structure. Here, using first-principles calculations, we investigate the adsorption of 3$d$ transition-metal atoms and dimers on monolayer $\alpha$-In$_2$Se$_3$ and identify Cr as a stable dimer configuration on both polar surfaces. Reversing the ferroelectric polarization drives a pronounced change in its magnetic coupling: the AFM state is favored  for one polarization, whereas the FM and AFM configurations become nearly degenerate for the opposite one. Analysis of the spin density, projected electronic states, and Wannier Hamiltonians reveals concomitant changes in the direct Cr–Cr hopping and in the interaction with the bridging Se atom. In particular, enhanced Cr–Se hybridization and Se spin polarization, together with a strong reduction of the direct two-center Cr–Cr hopping, point to a polarization-dependent balance between direct and substrate-mediated exchange contributions. These results show how ferroelectric polarization can tune magnetic interactions in an atomically defined 2D hybrid system.
\end{abstract}

\pacs{0}
\maketitle

\section{Introduction}\label{intro}
\indent

Controlling magnetic interactions at the atomic scale using an external electric field is a central goal in the development of low-dimensional spintronic devices. Ferroelectric materials are particularly attractive in this context because their switchable polarization can modify the local electrostatic environment, chemical bonding, and electronic structure of an adjacent magnetic system, providing a route to non-volatile control of magnetic states. Two-dimensional (2D) ferroelectrics are especially promising as substrates for this purpose, owing to their reduced dimensionality and atomically well-defined surfaces.

Among the available 2D ferroelectric materials, $\alpha$-In$_2$Se$_3$ is an especially attractive candidate because its robust ferroelectricity persists down to the monolayer limit ~\cite{Ding2017-teo, Zhou2017-exp,Wan2018-exp,Xue2018-exp}. In addition to its out-of-plane polarization, $\alpha$-In$_2$Se$_3$ exhibits an intercorrelated in-plane polarization, with polarization reversal involving a displacement of the central Se layer~\cite{Xiao2018}. This coupled structural and polar response gives rise to two inequivalent surface environments, making $\alpha$-In$_2$Se$_3$ particularly interesting for investigating how ferroelectric switching modifies the interaction between a substrate and adsorbed magnetic species. Its compatibility with other low-dimensional materials has also motivated growing interest in $\alpha$-In$_2$Se$_3$-based magnetic heterostructures, including systems involving FeI$_2$~\cite{Wei2019}, Cr$_2$Ge$_2$Te$_6$~\cite{Cheng2019}, Mn-decorated graphene~\cite{Picozzi2020}, and MnSe$_2$~\cite{Zang2024}.

Transition-metal adatoms and small clusters adsorbed on surfaces offer a means of probing magnetic interactions at the atomic scale. Within such nanostructures, magnetic coupling can result from competing mechanisms, including direct exchange between neighboring magnetic atoms and exchange mediated by the substrate. The balance between these contributions depends on factors such as interatomic distances, local bonding geometry, and the electronic structure of the surrounding atoms. While several $\alpha$-In$_2$Se$_3$-based magnetic heterostructures have been investigated~\cite{Wei2019,Cheng2019,Picozzi2020,Zang2024,Wang2021}, the influence of ferroelectric polarization on exchange interactions within adsorbed transition-metal clusters remains comparatively unexplored~\cite{Ting2020,Ju2021}.

In this work, we investigate how ferroelectric polarization affects the magnetic coupling between two Cr atoms forming a dimer adsorbed on a monolayer $\alpha$-In$_2$Se$_3$. Using first-principles calculations, we examine 3$d$ transition-metal species on the two polar surfaces of the ferroelectric substrate and identify Cr as the system in which a well-defined dimer is stabilized. 
We compare configurations in which the Cr magnetic moments are aligned parallel (ferromagnetic, FM) or antiparallel (antiferromagnetic, AFM).
We show that polarization reversal strongly modifies the balance between direct antiferromagnetic Cr–Cr exchange and a substrate-mediated contribution involving the bridging Se atom, reducing the robust AFM coupling for one polarization state to near-degeneracy between the FM and AFM configurations for the opposite polarization.

\section{Computational Details} \label{comp}
\indent

We performed spin-polarized electronic structure calculations based on density functional theory (DFT), using the projector augmented wave method (PAW) as implemented in the Vienna Ab initio Simulation Package (VASP)~\cite{VASP, PAW-VASP}. We explicitly treated the In (4d, 5s, 5p), Se (4s, 4p) and transition metals (TM) (3d, 4s) electrons as valence ones. The generalized gradient approximation (GGA) in the parametrization of Perdew, Burke, and Ernzerhof (PBE) was employed~\cite{PBE96}.

The DFT+$U$ approach, within Dudarev's formulation, was employed to account for the on-site Coulomb interactions among the localized 3$d$ electrons of the TM atoms. An effective Hubbard parameter of $U_{\rm eff}=4$ eV was adopted, within the range commonly used for 3$d$ transition metals~\cite{Hubbard2022, Hubbard2025}. In particular, this choice is consistent with previously reported values of approximately 2–4 eV for Cr adatoms and impurities in two-dimensional and metallic systems~\cite{Wang2018,Armitage2025}, as well as with linear-response estimates for Cr impurities in metallic environments~\cite{Hubbard2022}.

To verify the robustness of the results for the Cr adatom and dimer with respect to the choice of $U_{eff}$, we performed additional calculations using values ranging from 2 to 5 eV. The magnetic ground state of the Cr dimer remains AFM for one polarization state throughout this range, whereas for the opposite polarization state the energy difference between the FM and AFM configurations remains below 2 meV in magnitude.

The energy cutoff for the plane-wave basis was set at 500 eV. A total energy convergence better than $10^{–7}$ eV per unit cell was required, and the interatomic forces were minimized to 0.005 eV/Å for the structural relaxations. Long-range dispersion corrections were included through the DFT-D3 approach \cite{grimme2006semiempirical}. A dipole correction, as implemented in VASP, was applied along the direction perpendicular to the surface. To evaluate the electronic charge in individual atoms, we used the Bader approach  through the algorithm of Henkelman et al.\cite{HENKELMAN2006}.

The in-plane electric polarization was calculated using the conventional Berry-phase approach \cite{Resta1993}, whereas the out-of-plane component was determined by directly integrating the electronic charge density along the direction normal to the slab. To construct the polarization pathway within the Berry-phase formalism, eight intermediate images were generated using the nudged elastic band (NEB) method \cite{NEB2000}, connecting the ferroelectric (FE) ZB$'$ $\alpha$-phase with the paraelectric FCC phase~\cite{Ding2017-teo}. 

To model TM adatoms and dimers adsorbed on the $\alpha$-In$_2$Se$_3$ monolayer, we employed a 2$\times$2 surface supercell for the adatoms and 4$\times$4 and 4$\times$6 surface supercells for the dimers. A vacuum of 15~\AA\  was taken into account in the direction perpendicular to the surface to avoid fictitious interactions between periodic images. The first 2D Brillouin zones of the 2$\times$2, 4$\times$4 and 4$\times$6 supercells were sampled using 6$\times$6$\times$1, 3$\times$3$\times$1  and 3$\times$2$\times$1 $\Gamma$-centered Monkhorst-Pack grids, respectively. Convergence tests with respect to the k-point sampling are provided in Supplementary Material. 

To analyze the Cr–Se and Cr–Cr electronic couplings, we constructed tight-binding Hamiltonians from Wannier functions using Wannier90~\cite{wannier90_2014, wannier90_2020} interfaced with VASP. For each polarization state, the Hamiltonians were constructed for both spin channels of the FM configuration, using projections onto In-$s$, Se-$p$, Cr-$s$, and Cr-$d$ orbitals. The Cr-$s$ orbitals were included to describe Cr-$4s$-derived states close to the Fermi level. The Wannier subspace was selected using the disentanglement procedure~\cite{souza2001}, without subsequent maximal localization, in order to preserve as much as possible the atomic-orbital character of the initial projections. The resulting Hamiltonians accurately reproduce the DFT band structures within the energy window considered (see Supplementary Material for details).

The hopping blocks between the bridging Se and each Cr atom, as well as between the two Cr atoms, were fitted by least squares using two-center Slater–Koster parameters~\cite{slater1954}. The quality of each fit was assessed from the fraction of the squared norm of the corresponding hopping block reproduced by the two-center model. We define $\Delta_{pd}$ as the difference between the average on-site energies of the Cr-$d$ and bridging Se-$p$ Wannier functions.

\section{Results and discussion}\label{results}
\indent
Indium selenide is a layered semiconductor material that exhibits several structural phases. Among them, the $\alpha$-phase has been predicted and experimentally demonstrated to possess robust ferroelectricity at room temperature, both in its few-layer and monolayer forms~\cite{Ding2017-teo,Zhou2017-exp,Wan2018-exp,Xue2018-exp}. The $\alpha$-phase crystallizes in a hexagonal (2H) structure consisting of five covalently bonded atomic layers in the sequence Se–In–Se–In–Se, with two distinct coordination environments for the In atoms: tetrahedral, In(t), and octahedral, In(o). Correspondingly, the two outer Se atomic planes are denoted Se(t) and Se(o), according to their association with In(t) and In(o), respectively, while Se(m) denotes the central Se plane. In the bulk and multilayered forms, adjacent quintuple layers (QLs) are held together by weak van der Waals interactions. 

\begin{figure}
\raggedleft
\includegraphics[width=1.\textwidth]{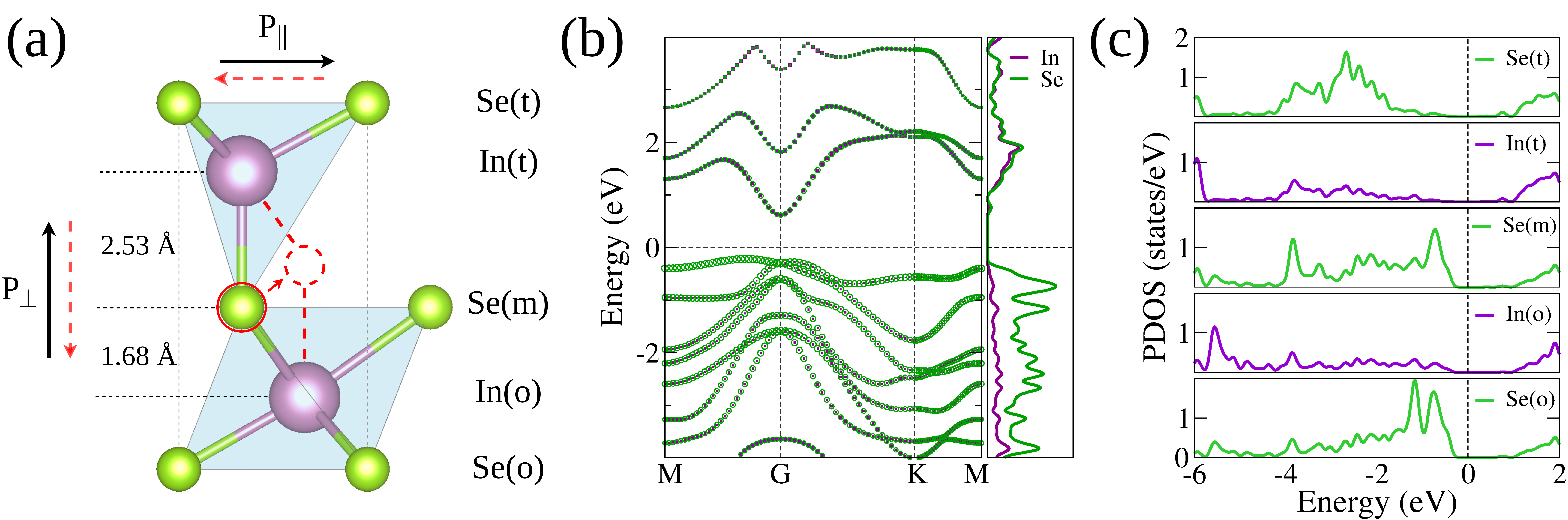}
\caption{(a) Schematic side view of the relaxed pristine $\alpha$-In$_2$Se$_3$ quintuple layer, showing the two inequivalent In coordination environments: tetrahedral, In(t), and octahedral, In(o) and the corresponding Se atoms: Se(t), Se(m) (central), and Se(o). In-plane ($P_\parallel$) and out-of-plane ($P_\perp$) polarization are intercorrelated in this material. The red arrow and open circle illustrate the displacement of the central Se(m) atom associated with ferroelectric switching between the P$\uparrow$ and P$\downarrow$ polarization states. The purple (green) spheres represent In (Se) atoms. (b) Corresponding electronic band structure and partial density of states, with the orbital character of In (Se) indicated in violet (green). The Fermi level is set to 0 eV. (c) Layer-resolved partial density of states projected onto the Se and In atoms.
}
\label{fig:In2Se3-pristine}
\end{figure}

We first calculated the structural and electronic properties of the fully relaxed pristine monolayer. 
The relaxed QL structure is shown in Fig.~\ref{fig:In2Se3-pristine} (a). The optimized in-plane lattice parameter is 4.04~\AA, in good agreement with the experimentally reported value of approximately 4.03~\AA\,\ for bulk $\alpha$-In$_2$Se$_3$~\cite{kuppers2018, Balakrishnan_2018}. The remaining structural parameters are reported in Table~\ref{tab:SI-struct}. The calculated in-plane polarization difference is $\Delta P_{\parallel}$=2.35 e\AA, oriented at an angle of $30^\circ$ with respect to the horizontal axis, while the out-of-plane component is $\Delta P_{\perp}$ = 0.095 e\AA. Both values are in excellent agreement with previously reported results~\cite{Ding2017-teo}.

The corresponding band structure, shown in Fig.~\ref{fig:In2Se3-pristine} (b), confirms the semiconducting character of the In$_2$Se$_3$ monolayer, with an indirect PBE bandgap of 0.83~eV. As expected, this value is smaller than the experimental bandgap of approximately 1.3 eV, owing to the well-known band-gap underestimation within the PBE approximation~\cite{Cui2018-exp-gap}.

The projected density of states (PDOS) in Fig.~\ref{fig:In2Se3-pristine}(c) reveals a pronounced asymmetry between the two surface Se atoms, with the Se(o)-$p$ states lying closer to the Fermi level than the corresponding Se(t)-$p$ states. In the following, we denote the surface terminated by Se(t) as the $P\uparrow$ surface, and the corresponding one terminated by Se(o) as the $P\downarrow$ surface, according to the direction of the ferroelectric polarization. 

\begin{figure}
\centering
\includegraphics[width=1\textwidth]{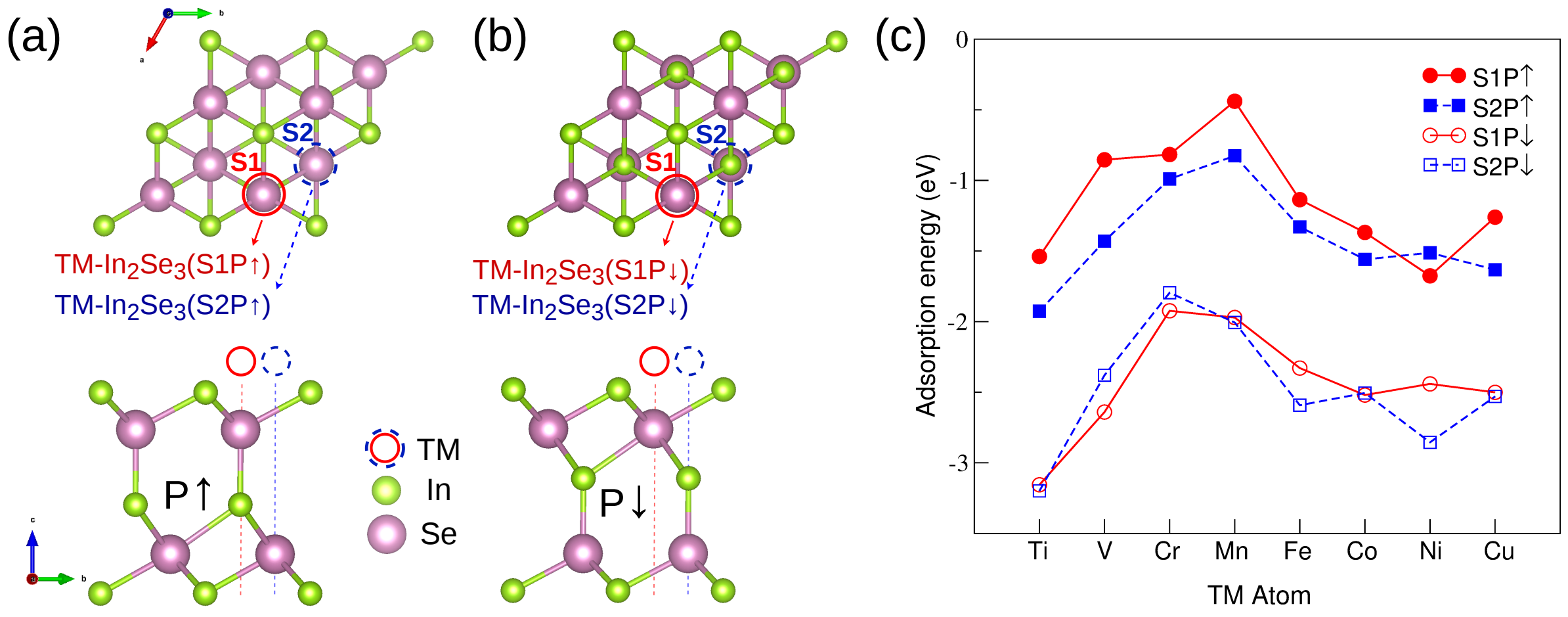}
\caption{Top and side views of the In$_2$Se$_3$ (2$\times$2) supercell for the (a) P$\uparrow$ and (b) P$\downarrow$ polarization directions, showing the two adsorption sites considered for the TM adatoms: S1 site indicated by a red circle and S2 site indicated by a blue circle. In the side views, the black arrow indicates the direction the out-of-plane polarization. The purple and green spheres represent In and Se atoms, respectively. (c) Adsorption energies of the TM adatoms at the S1 and S2 sites of In$_2$Se$_3$ for the two ferroelectric polarization states, P$\uparrow$ and P${\downarrow}$. Filled and open symbols correspond to P${\uparrow}$ and P${\downarrow}$, respectively, while circles and squares correspond to the S1 and S2 sites, respectively.}
\label{fig:Eads}
\end{figure}

To investigate how ferroelectric polarization affects transition-metal adsorption, we considered two adsorption sites on each surface of the In$_2$Se$_3$ QL. Both sites correspond to threefold hollow sites of the surface Se layer: S1 is located above an In atom in the second layer, whereas S2 is located above an In atom in the fourth layer, as indicated by the red and blue circles in Figs.~\ref{fig:Eads}(a,b).

The adsorption energy, $E_{ads}$, is defined as:
\begin{equation}
    E_{ads}=E_{TM@In_2Se_3} - E_{In_2Se_3} - E_{TM}
\end{equation}
where $E_{TM@In_2Se_3}$ and $E_{In_2Se_3}$ are the total energies of the In$_2$Se$_3$ substrate with and without TM adatoms, respectively. $E_{TM}$ is the total energy of the isolated TM atom, computed in a 10.0\AA$ \,\times$ 10.1\AA \,$\times$ 10.2\AA \, box. 

The calculated adsorption energies, shown in Fig.~\ref{fig:Eads}(c), range from about $-3.0$ to $-0.5$ eV, indicating that adsorption is energetically favorable for all the studied transition-metal adatoms. These values are consistent with chemisorption for most of the systems.

For all transition-metal adatoms and at both adsorption sites, adsorption is systematically more favorable on the P$\downarrow$ surface than on the P$\uparrow$ one. 
This polarization dependence reflects the distinct structural and electronic environments of the two surfaces.
In particular, the In–Se(o) bonds at the P$\downarrow$ surface are longer than the corresponding In–Se(t) bonds at P$\uparrow$ (Table~\ref{tab:SI-struct}).
This structural difference is accompanied by a lower Bader charge of the Se(o) atoms compared with Se(t) (Table~\ref{tab:SI-bader}) and by Se(o)-$p$ states lying closer to the Fermi level than the corresponding Se(t)-$p$ states (Fig.~\ref{fig:In2Se3-pristine} (c)). These features suggest that the Se(o)-$p$ states are more readily available for the interaction with the adsorbed transition-metal orbitals, contributing to the stronger adsorption observed on the P$\downarrow$ surface.

The local structural distortions induced by adsorption are more pronounced at site S1 than at site S2. At S1, the adatom is fourfold-coordinated to three surface Se atoms and one subsurface In atom, which is significantly displaced downward from its ideal position upon relaxation. In contrast, adsorption at S2 involves three surface Se atoms and induces smaller, more localized distortions (Fig. S2). Similar relaxation patterns are obtained for all studied TM adatoms. We therefore selected S2 for the subsequent calculations because it better preserves the underlying QL structure and its polar distortion.

With this choice, we study pairs of 3$d$ transition-metal adatoms placed at two nearest-neighbor S2 sites on each polar termination, hereafter referred to as dimers. Three geometrically distinct dimer orientations with respect to the in-plane polarization direction are possible, as shown in Fig.~\ref{fig:SIdimer-orientation}. We compared two of these orientations and found no appreciable differences in their structural, electronic, or magnetic properties. For the remainder of this work, we focus on orientation O2.

\begin{figure}[h!]
\centering
\includegraphics[width=.7\textwidth]{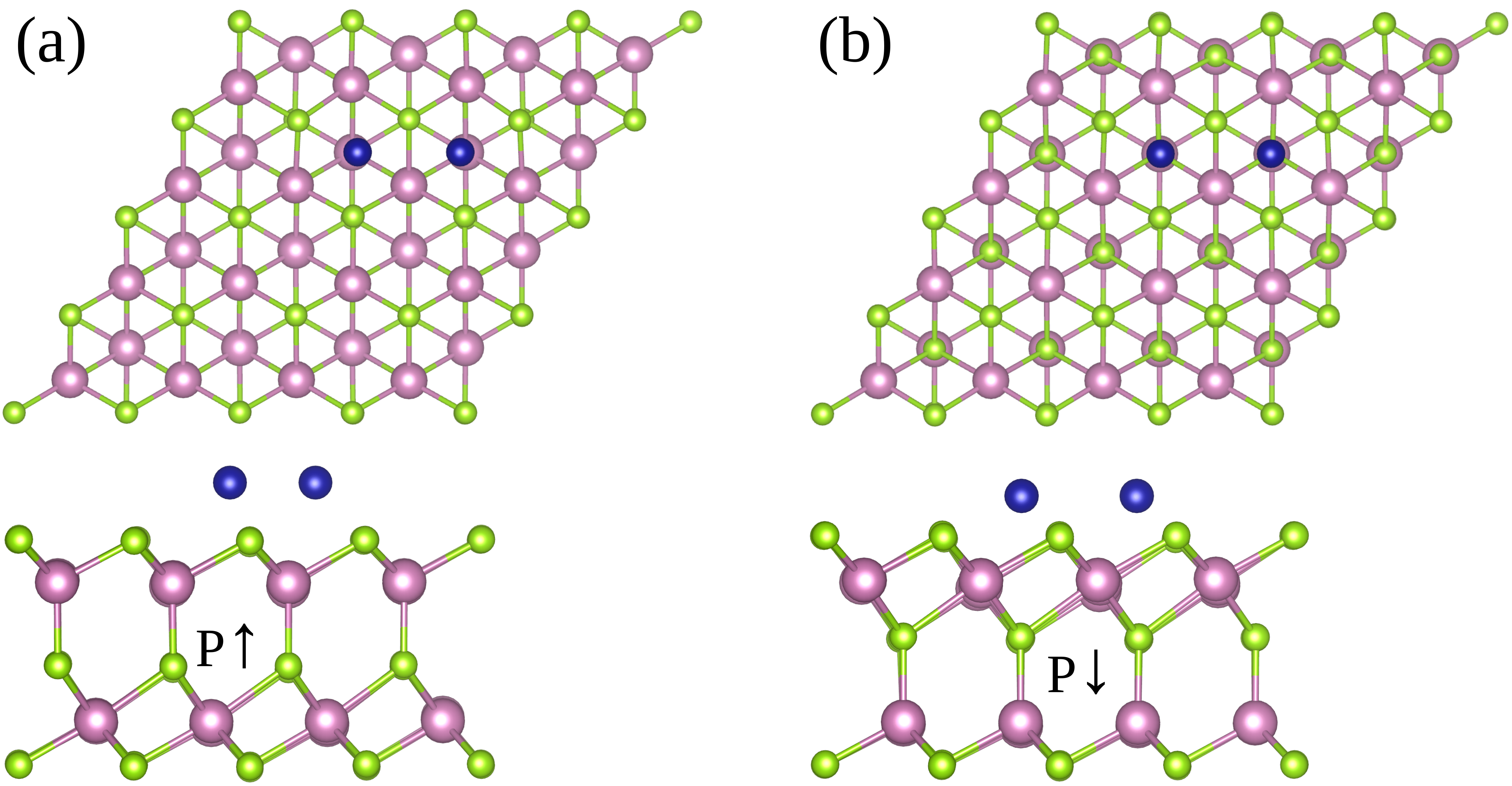}
\caption{Top and side views of the relaxed Cr dimer adsorbed at the S2 site on monolayer In$_2$Se$_3$ for (a) P$\uparrow$ and (b) P$\downarrow$. Cr atoms are shown in blue.} 
\label{fig:Cr-dimer-4x4}
\end{figure}

After structural relaxation, the Cr dimer is the only one among all the 3$d$ dimers examined on both surfaces that retains its initial configuration without significantly distorting the underlying substrate (Fig.~\ref{fig:Cr-dimer-4x4}). In contrast, the other transition-metal dimers undergo substantial structural rearrangements involving the adatoms and/or the underlying QL, preventing a direct comparison of their magnetic properties within a common dimer geometry. The structural stability of the Cr dimer therefore makes it a particularly suitable system for investigating the magnetic coupling between two adatoms and its dependence on ferroelectric polarization. The relaxed Cr dimer structures corresponding to the two polarization states are shown in Fig.~\ref{fig:Cr-dimer-4x4}. 

Bader charge analysis shows that all the TM adatoms transfer electronic charge to the In$_2$Se$_3$ substrate, shifting the Fermi level toward the conduction-band states and thus producing local electron doping of the QL (Table~\ref{tab:SIproperties_TM}). The charge transferred, $\Delta Q$, defined as the difference between the nominal valence charge of the isolated TM atom and its Bader charge, is generally larger on P$\downarrow$ than on P$\uparrow$, with Co showing negligible polarization dependence. For most adatoms, this difference is accompanied by shorter TM–Se(surface) distances on P$\downarrow$ (Table~\ref{tab:SIproperties_TM}), indicating that polarization reversal modifies the local TM–substrate interaction.

Among the TM adatoms studied, Cr shows the largest TM–Se(surface) distance for P$\uparrow$ and, together with Ti, the largest distance for P$\downarrow$. Unlike Ti, 
Cr combines these relatively long TM–Se distances with a comparatively weak adsorption energy (Fig.~\ref{fig:Eads}),
indicating that Cr interacts less strongly with the substrate than most of the other adatoms considered. This provides a natural explanation for why the Cr adatoms can form a dimer while inducing only minor structural rearrangements of the underlying QL.

\begin{figure}
\centering
\includegraphics[width=0.8\textwidth]{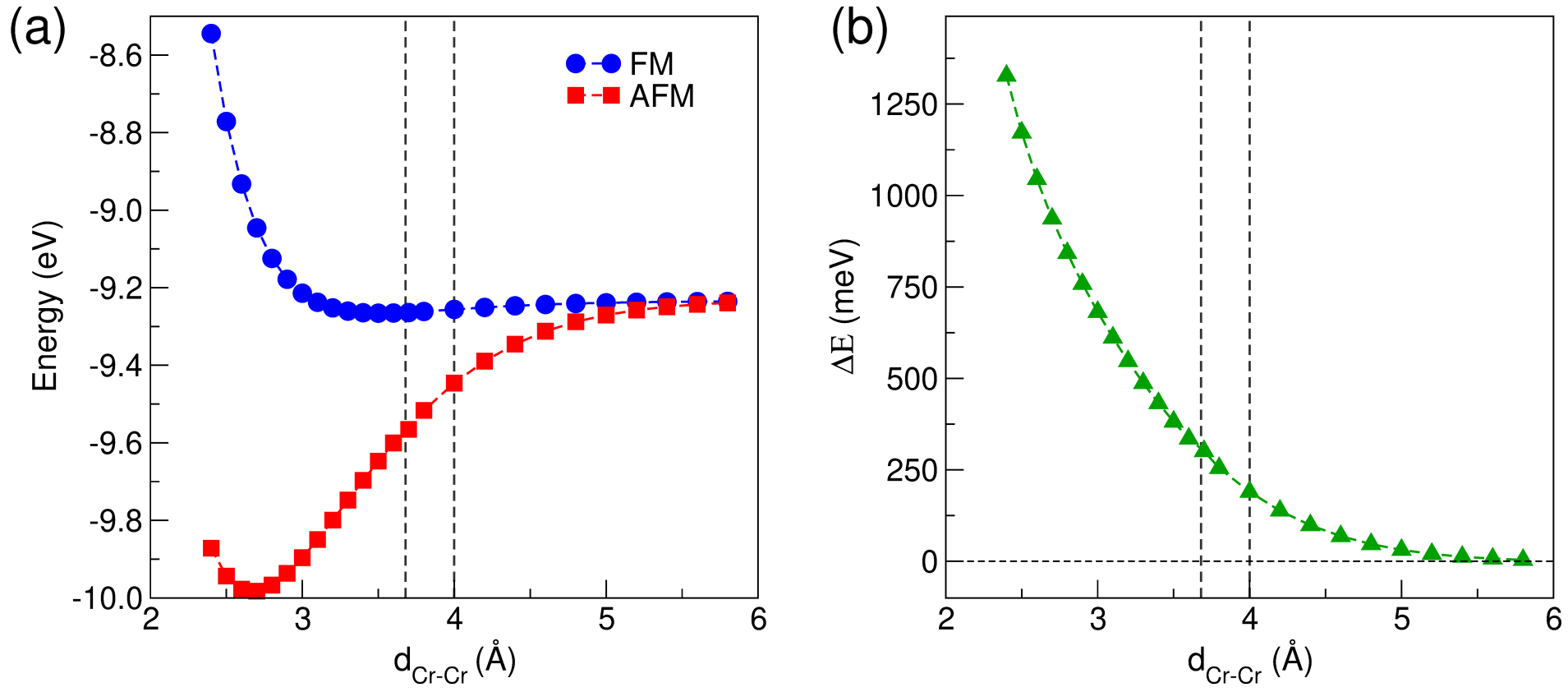}
\caption{(a) Total energy of the isolated Cr dimer in vacuum, in the FM (blue, circles) and AFM (red, squares) configurations, as a function of the Cr-Cr separation, $d_{Cr-Cr}$. (b) Corresponding energy difference $\Delta E = E_{FM}-E_{AFM}$ as a function of distance. Positive values indicate that the AFM configuration is lower in energy. The vertical dashed lines mark the Cr–Cr separations obtained for the adsorbed dimer in the two polarization states.}
\label{fig:Cr-dimer-isolated}
\end{figure}

To establish a qualitative reference for the strength and distance dependence of the direct Cr–Cr exchange interaction, we first considered an isolated Cr dimer in vacuum, using the same exchange-correlation functional and Hubbard
correction ($U_{eff}=4$~eV) as the adsorbed systems. We compared the FM and AFM alignment of the Cr magnetic moments and  
define $\Delta E=E_{FM}-E_{AFM}$, such that positive (negative) values favor AFM (FM) ordering. Figure~\ref{fig:Cr-dimer-isolated}(a) shows the total energies of both configurations as a function of the Cr–Cr separation. The AFM configuration is the ground state at all distances considered, with $\Delta E$ decreasing monotonically from several hundred meV near the equilibrium separation ($\approx$ 2.7~\AA) to only a few meV beyond $\sim$5.5~\AA\ (Fig.~\ref{fig:Cr-dimer-isolated}(b)). This behavior highlights the rapid decay of the direct Cr–Cr exchange interaction with increasing separation. 
Hereafter, we denote by $\Delta E_{\rm iso}$ the energy difference of the isolated dimer at a given Cr–Cr separation. 

The equilibrium Cr–Cr distances in the surface-supported dimers, 3.68~\AA\ for P$\uparrow$ and 3.97~\AA\ for P$\downarrow$, are substantially larger than that of the isolated dimer, suggesting that the adsorbed configurations are stabilized primarily by interactions with the substrate rather than by strong direct Cr–Cr bonding. Nevertheless, the absence of a strong chemical bond does not preclude a significant direct Cr–Cr magnetic exchange interaction at these separations,  which may provide an AFM contribution to the overall magnetic coupling, as discussed below. Because adsorption also modifies the charge state and electronic configuration of the Cr atoms, the magnetic coupling in the supported dimer cannot be quantitatively decomposed into independent direct Cr–Cr and substrate-mediated contributions on the basis of the isolated-dimer results alone.

The magnetic coupling and relevant structural parameters of the surface-supported Cr dimers are summarized in Table~\ref{tab:summary}. To verify that the results are not affected by interactions between periodic images, we performed calculations using both $4\times4$ and $4\times6$ surface supercells, with the dimer oriented along the direction extended in the 4 × 6 supercell. Both supercell sizes yield the same qualitative picture: a robust AFM ground state for P$\uparrow$, with $\Delta$E of the order of 80–90 meV, and FM–AFM degeneracy for P$\downarrow$, with $|\Delta E|$ below 1 meV (Table~\ref{tab:SI-4x6}). Although a small quantitative difference is observed for P$\uparrow$, the strong polarization dependence of the magnetic coupling remains unchanged, supporting the use of the 4 $\times$ 4 supercell for the present analysis. The results reported in Table~\ref{tab:summary} were obtained using the last supercell.

\begin{table}[htbp]
\centering
\caption{Structural parameters and magnetic energy differences for the Cr dimer on the two polar surfaces of $\alpha$-In$_2$Se$_3$. $\Delta E = E_{FM}-E_{AFM}$; positive values indicate an AFM ground state. The isolated-dimer values, $\Delta E_{iso}$, are calculated at the Cr–Cr distances obtained after relaxation on the corresponding surface.}
\label{tab:summary}
\small
\setlength{\tabcolsep}{3pt}
\begin{tabular}{cc ccc cc}
\hline
Pol. & \begin{tabular}{@{}c@{}}Surface\\term.\end{tabular} & \begin{tabular}{@{}c@{}}$d_{\mathrm{Cr-Cr}}$\\(\AA)\end{tabular} & \begin{tabular}{@{}c@{}}$d_{\mathrm{Cr-Se}}$\\(\AA)\end{tabular} & \begin{tabular}{@{}c@{}}$\angle$Cr-Se-Cr\\(deg)\end{tabular} & \begin{tabular}{@{}c@{}}$\Delta E_{\mathrm{iso}}$\\(meV)\end{tabular} & \begin{tabular}{@{}c@{}}$\Delta E$\\(meV)\end{tabular} \\
\hline
P$\uparrow$ & Se(t) & 3.68 & 2.73 & 84.7 & +308 & 89.9 \\
P$\downarrow$ & Se(o) & 3.97 & 2.63 & 97.9 & +199 & -0.2 \\
\hline
\end{tabular}
\end{table}

For P$\uparrow$, the Cr dimer remains AFM, with $\Delta E = 89.9$ meV, substantially smaller than 308 meV obtained for the isolated dimer at the same Cr–Cr distance. 
For P$\downarrow$, the FM and AFM configurations are degenerate within the numerical precision of our calculations, with $|\Delta E|$ below 1 meV in both supercells. The sign of such a small energy difference is not physically meaningful; the relevant result is the complete suppression of the magnetic coupling relative to P$\uparrow$.
These results constitute the central finding of this work: polarization reversal tunes the Cr dimer from a robust AFM state to a near-degenerate FM–AFM state.

The substantial reduction in the magnetic energy difference upon adsorption, together with its strong dependence on the polarization state, indicates that the magnetic coupling cannot be explained exclusively in terms of direct Cr–Cr exchange.
Although the increase in Cr–Cr separation upon polarization reversal also weakens the magnetic coupling of the isolated dimer, the corresponding change is much less pronounced than that observed for the adsorbed dimer. 
Assuming a qualitatively similar distance dependence of the direct exchange in the adsorbed system, this comparison suggests that the change in Cr–Cr separation alone may not account for the pronounced reduction in the magnetic energy difference upon polarization reversal. 

We associate this additional effect with a possible substrate-mediated exchange contribution involving the bridging Se atom, hereafter denoted as Se$_{br}$, which competes with the direct Cr–Cr exchange interaction.
The bridging atom corresponds to Se(t) for P$\uparrow$ and Se(o) for P$\downarrow$, reflecting the polarization-dependent structural and electronic asymmetry previously discussed for the pristine substrate. The substrate-mediated magnetic coupling is therefore expected to depend on both the local Cr–Se$_{br}$–Cr geometry and the electronic structure of the bridging Se atom, which are modified upon polarization reversal.
 
Structural relaxation provides a first indication of how the substrate-mediated interaction changes when the polarization is switched. 
The Cr–Se distance to the bridging atom becomes shorter for P$\downarrow$, whereas the Cr–Se$_{br}$–Cr angle remains close to 90$^\circ$ in both polarization states. 
The shorter Cr–Se distance suggests an enhanced interaction between Cr and the bridging Se atom in the P$\downarrow$ configuration. The near-90$^\circ$ Cr–Se–Cr geometry is compatible with a superexchange pathway within the Goodenough–Kanamori–Anderson framework, in which a ferromagnetic contribution may arise for approximately 90$^\circ$ metal–ligand–metal bond angles. However, the P$\downarrow$ configuration, which exhibits a much smaller net AFM exchange energy, has a slightly larger deviation from 90$^\circ$. Thus, the bond angle alone cannot account for the stronger FM contribution in P$\downarrow$, suggesting that the enhanced Cr–Se$_{br}$ interaction also plays an important role.

The local magnetic moments within the PAW spheres offer an additional perspective on the magnetic response of the bridging Se atom.
In the FM configuration, the Cr atoms retain an electronic configuration dominated by occupied majority-spin $d$ states and essentially empty minority-spin $d$ states for both polarization directions, while the bridging Se atom develops a finite magnetic moment antiparallel to the Cr moments. The magnitude of the Se$_{br}$ moment increases from 0.03 to 0.07 $\mu_B$ upon switching from P$\uparrow$ to P$\downarrow$, whereas the Cr moments decrease only slightly. 
This increase indicates a stronger spin polarization of the bridging Se atom in the P$\downarrow$ configuration, supporting its enhanced involvement in the magnetic response of the Cr dimer.

\begin{figure}
\centering
\includegraphics[width=.9\columnwidth]{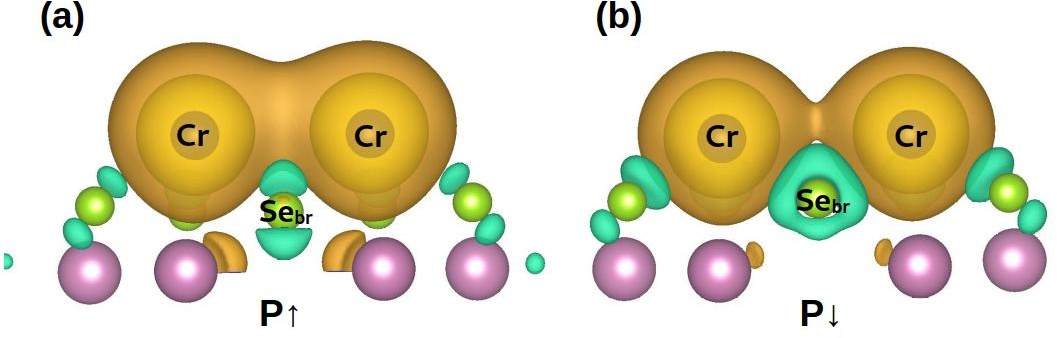}
\caption{Isosurfaces of the spin density ($\rho_\uparrow-\rho_\downarrow$) for the Cr dimer in the FM configuration, on the (a) P$\uparrow$ and (b) P$\downarrow$ surfaces, at a fixed isosurface value of 0.0018~$\mu_B$~\AA$^{-3}$. Yellow (cyan) surfaces indicate positive (negative) spin density. In both cases, the spin density is concentrated on the Cr atoms and extends onto the bridging Se atom - Se(t) on P$\uparrow$, Se(o) on P$\downarrow$ - with opposite sign relative to the Cr spin polarization.
}
\label{fig:Cr-dimer-spin}
\end{figure}

The real-space spin density provides complementary information on the spatial distribution of this polarization. Fig.~\ref{fig:Cr-dimer-spin} shows the spin-density distribution for the FM Cr dimers at the same isosurface value of 0.0018 $\mu_B$ \AA$^{-3}$. In both polarization states, the spin density is concentrated on the Cr atoms and extends onto the bridging Se atom, with opposite sign relative to the Cr spin polarization. The spin-density distribution around the Se$_{br}$ is more pronounced for P$\downarrow$, revealing a stronger and more spatially extended polarization of the bridging ligand. This spatial redistribution provides further evidence that the bridging Se participates in the magnetic coupling between the two Cr atoms. The spin-density profiles along the Cr–Se$_br$–Cr bonding path for both FM and AFM configurations are shown in the Supplementary Information (Fig.~\ref{fig:Cr-dimer-perfil-spin}).

\begin{figure}
\centering
\includegraphics[width=.8\columnwidth]{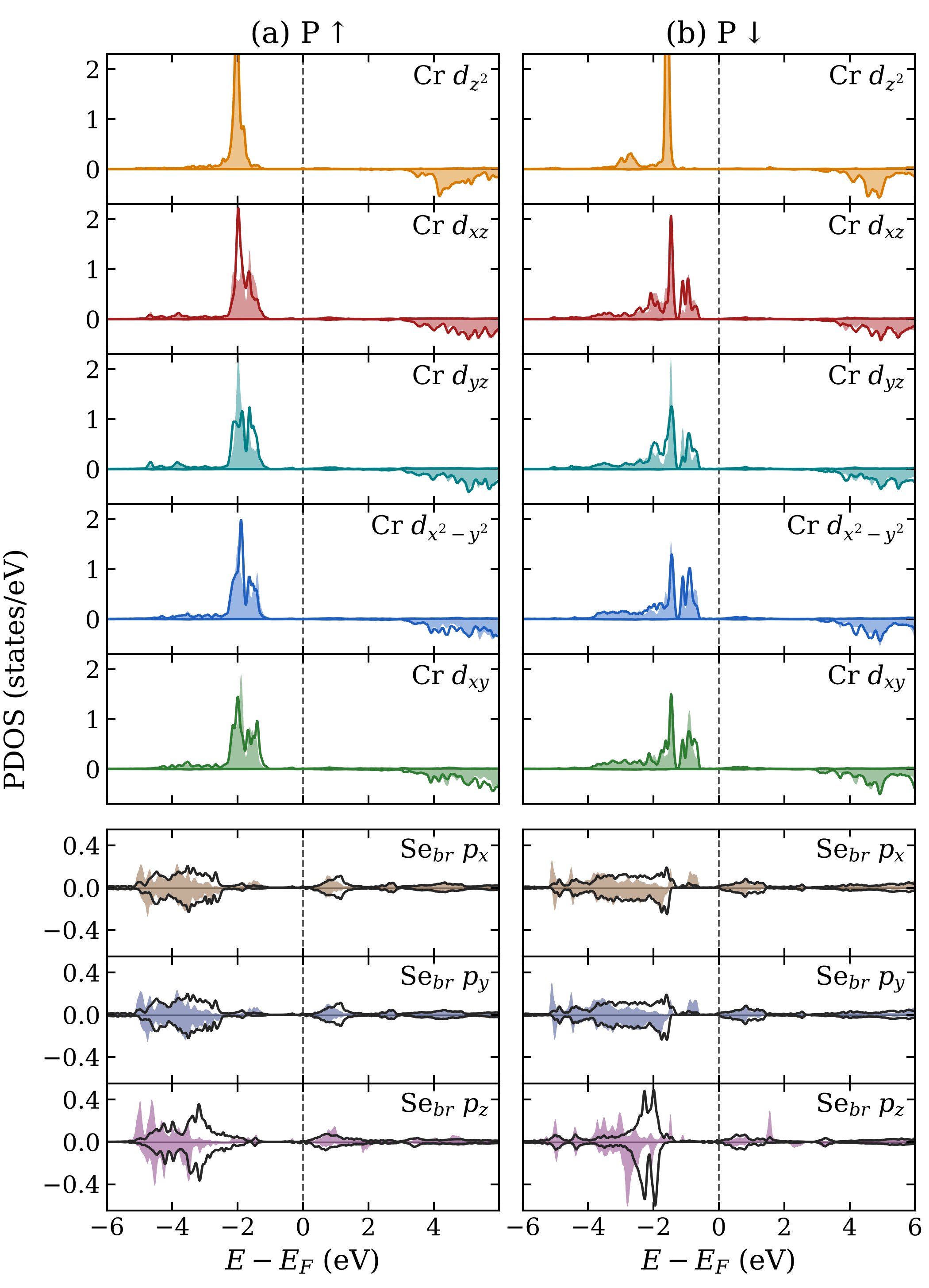}
\caption{\label{fig:pdos}Spin-resolved projected density of states of the Cr
dimer in the FM configuration on the (a) P$\uparrow$ and (b) P$\downarrow$
surfaces. Upper panels: Cr-$d$ orbitals (shaded: Cr1; line: Cr2). Lower
panels: $p$ orbitals of the bridging Se atom (shaded) and average over the
other surface Se atoms, excluding Se$_{br}$ (line). Positive (negative) values
correspond to majority (minority) spin. Orbitals are referred to the Cartesian
axes of the supercell, with $z$ normal to the surface. The energy zero is the
Fermi level (dashed line).}
\label{fig:PDOS}
\end{figure}  

The projected density of states (PDOS) provides further insight into the electronic origin of this Se-mediated interaction. In both polarization regimes, the Cr-$d$ states are dominated by majority-spin contributions, consistent with the nearly $d^5$ configuration of the Cr atoms. The PDOS of the bridging Se atom shows greater spectral overlap with the Cr-$d$ states around -2 eV for P$\downarrow$ (Fig.\ref{fig:PDOS}), suggesting enhanced Cr–Se$_{br}$ electronic hybridization upon polarization reversal. To quantify the shared orbital character of these states, we evaluated a mixing indicator, $M$, based on the product of the Cr-$d$ and Se$_{br}$-$p$ projection weights of the occupied majority-spin Kohn–Sham states (see Supplementary Information for details). The value of $M$ increases from 0.039 for P$\uparrow$ to 0.114 for P$\downarrow$, confirming the larger shared Cr-$d$/Se$_{br}$-$p$ character of the occupied states in the latter configuration.

To further characterize the electronic couplings associated with this orbital mixing, we analyzed the Wannier Hamiltonians and fitted the Cr-Se$_{br}$ and Cr-Cr hopping blocks using two-center Slater-Koster parameters (Table~\ref{tab:tb}). The Cr-Se$_{br}$ hopping blocks are very well reproduced by the two-center model in both 
polarization states.
Upon polarization reversal, the Cr-Se$_{br}$ hopping parameters increase in magnitude, consistent with the reduction in the Cr-Se$_{br}$ bond length. In particular, Harrison’s (d$^{-7/2}$) scaling for p-d hoppings~\cite{Harrison1980} predicts an increase of about 14\%, in good agreement with the calculated values. 

In addition, the energy separation $\Delta_{pd}$ between the Cr-$d$ and bridging Se-$p$ levels 
decreases by about 0.6~eV, reflecting the upward shift of the Se(o)-$p$ states discussed for the pristine
surface (Fig.\ref{fig:In2Se3-pristine}(c)). The combined effect of the enhanced hopping amplitudes and reduced energy separation favors stronger Cr-$d$/Se$_{br}$-$p$ hybridization for P$\downarrow$, consistent with the larger mixing indicator  $M$ obtained for this polarization.

\begin{table}[htbp]
\caption{\label{tab:tb}Tight-binding parameters of the majority-spin channel obtained from the Wannier Hamiltonians. $V_{pd\sigma}$, $V_{pd\pi}$ and $V_{sp\sigma}$ are the Slater-Koster parameters of the bond between the bridging Se, Se$_{br}$ and each Cr atom; $\Delta_{pd}$ is the energy separation between the average on-site energies of the Cr-$d$ and Se$_{br}$-$p$ Wannier functions; $V_{dd\sigma}$ and $V_{dd\pi}$ are the two-center hoppings between the two Cr atoms, obtained from a fit of the Cr–Cr block with $V_{dd\sigma}$, $V_{dd\pi}$ and $V_{dd\delta}$; $V_{dd\delta}$ is negligible ($|V_{dd\delta}| < $ 10 meV) in both polarization states and is not listed. The last two rows give the fraction of the squared norm of the Cr–Cr hopping block described by this two-center form and the norm of the remaining part. 
}
\begin{center}
\begin{tabular}{lccc}
\hline
 & P$\uparrow$ & P$\downarrow$ & P$\downarrow$/P$\uparrow$ \\
\hline
$d_{\rm Cr-Se_{br}}$ (\AA)            & 2.732   & 2.633   & 0.96 \\
$d_{\rm Cr-Cr}$ (\AA)                 & 3.679   & 3.972   & 1.08 \\
$V_{pd\sigma}$ (meV)                  & \phantom{-}-853  & \phantom{-}-958  & 1.12 \\
$V_{pd\pi}$ (meV)                     & \phantom{--}458   & \phantom{--}547   & 1.19 \\
$V_{sp\sigma}$ (meV)                  & -1159 & -1372 & 1.18 \\
$\Delta_{pd}$ (eV)                    & \phantom{-}1.83    & \phantom{-}1.25    & 0.68 \\
$V_{dd\sigma}$ (meV)                  & \phantom{-}-195  & \phantom{--}-83   & 0.43 \\
$V_{dd\pi}$ (meV)                     & \phantom{---}55    & \phantom{---}26    & 0.47 \\
Two-center fraction of Cr-Cr block    & \phantom{---}68\%    & \phantom{---}20\%    & --   \\
Norm of non-two-center part (meV)     & 145     & 182     & 1.26 \\
\hline
\end{tabular}
\end{center}
\end{table}

The Cr-Cr coupling exhibits a markedly different behavior. Although the Cr-Cr distance increases only moderately upon polarization reversal, the direct two-center hoppings decrease much more strongly than expected from Harrison’s $d^{-5}$ scaling. In particular, their reduction is substantially larger than the 32\% decrease predicted from the change in bond length alone. Moreover, the two-center approximation captures a much smaller fraction of the Cr-Cr hopping block for P$\downarrow$ than for  P$\uparrow$. Importantly, this change is accompanied by an increase in the norm of the residual hopping block, indicating that the contribution not captured by the two-center description also grows in absolute magnitude. Thus, the change in the Cr-Cr coupling cannot be understood solely in terms of the increased Cr-Cr distance. Instead, it points to a substantially enhanced role of environment-dependent or multicenter hopping processes in P$\downarrow$, potentially involving the substrate atoms shared by the two Cr atoms.

The near 90$^\circ$ Cr–Se$_{br}$–Cr geometry provides a structural environment compatible with a ligand-mediated exchange pathway within the Goodenough–Kanamori–Anderson framework. In this picture, coupling through approximately orthogonal ligand orbitals can give rise to a ferromagnetic contribution through Hund's coupling on the ligand. The enhanced Se$_{br}$ spin polarization and Cr–Se$_{br}$ spectral overlap observed for P$\downarrow$ are consistent with a stronger contribution of this substrate-mediated exchange channel.

 At the same time, the increased Cr–Cr separation and the strong suppression of the direct two-center Cr–Cr hoppings point to a weakening of the direct AFM interaction. Taken together, these results support an interpretation in which polarization reversal modifies the balance between direct AFM and ligand-mediated FM exchange contributions, leading to the near-degeneracy of the FM and AFM configurations for P$\downarrow$.

\section{Conclusions}

In summary, our first-principles calculations demonstrate that ferroelectric polarization provides an effective means of tuning the magnetic coupling of a Cr dimer adsorbed on $\alpha$-In$_2$Se$_3$. Polarization reversal changes the magnetic energy difference from a robust AFM preference for P$\uparrow$ to a near-degeneracy of the FM and AFM configurations for $P\downarrow$. This behavior is accompanied by substantial changes in the Cr–Cr and Cr–Se couplings, enhanced Cr-$d$/Se-$p$ hybridization, and increased spin polarization of the bridging Se atom. These results support a polarization-dependent balance between direct Cr–Cr exchange and substrate-mediated interactions, although their individual contributions cannot be quantitatively separated within the present analysis. Our findings highlight the role of ferroelectric substrates in controlling magnetic interactions at the atomic scale and suggest a route toward electrically tunable magnetic states in low-dimensional hybrid systems.

\section*{Acknowledgments}
This work was partially supported by PICT-2019-2128 of the ANPCyT, Argentina. 

\section*{Bibliography}
\bibliographystyle{iopart-num}
\bibliography{bib-vdw.bib}

@article{Resta1993,
  title = {Towards a quantum theory of polarization in ferroelectrics: The case of ${\mathrm{KNbO}}_{3}$},
  author = {Resta, R. and Posternak, M. and Baldereschi, A.},
  journal = {Phys. Rev. Lett.},
  volume = {70},
  issue = {7},
  pages = {1010--1013},
  numpages = {0},
  year = {1993},
  month = {Feb},
  publisher = {American Physical Society},
}

@ARTICLE{VASP,
author = {G. Kresse and J. Furthm{\"u}ller},
title = {Efficiency of ab-initio total energy calculations for metals and semiconductors},
journal = {Phys. Rev. B},
volume = {54},
pages = {11169},
year = {1996},
}

@ARTICLE{PBE96,
author = {J. P. Perdew and S. Burke and M. Ernzerhof},
title = {Generalized Gradient Approximation Made Simple},
journal = {Phys. Rev. Lett.},
volume = {77},
pages = {3865},
year = {1996},
}

@ARTICLE{PAW-VASP,
author = {G. Kresse and D. Joubert},
title = {From ultrasoft pseudopotentials to the projector augmented-wave method},
journal = {Phys. Rev. B},
volume = {59},
pages = {1758},
year = {1999},
}

@article{NEB2000,
    author = {Henkelman, Graeme and Uberuaga, Blas P. and Jónsson, Hannes},
    title = {A climbing image nudged elastic band method for finding saddle points and minimum energy paths},
    journal = {The Journal of Chemical Physics},
    volume = {113},
    number = {22},
    pages = {9901-9904},
    year = {2000},
    month = {12},
}

@article{grimme2006semiempirical,
  title={Semiempirical GGA-type density functional constructed with a long-range dispersion correction},
  author={Grimme, Stefan},
  journal={Journal of computational chemistry},
  volume={27},
  number={15},
  pages={1787--1799},
  year={2006},
  publisher={Wiley Online Library}
}

@article{HENKELMAN2006,
title = {A fast and robust algorithm for Bader decomposition of charge density},
journal = {Computational Materials Science},
volume = {36},
number = {3},
pages = {354-360},
year = {2006},
author = {Graeme Henkelman and Andri Arnaldsson and Hannes Jónsson}
}

@article{Ding2017-teo,
author = {Wenjun Ding and Jianbao Zhu and Zhe Wang and Yanfei Gao and Di Xiao and Yi Gu  and Zhenyu Zhang and Wenguang Zhu },
title = {Prediction of intrinsic two-dimensional ferroelectrics in In2Se3 and other III2-VI3 van der Waals materials},
journal = {Nature Communications} ,
volume = {8},
pages = {14956},
year = {2017}
}

@article{Zhou2017-exp,
author = {Yu Zhou and Di Wu and Yihan Zhu and Yujin Cho and Qing He and Xiao Yang  and Kevin Herrera and Zhaodong Chu and Yu Han and Michael C. Downer and Hailin Peng and Keji Lai},
title = {Out-of-Plane Piezoelectricity and Ferroelectricity in Layered α-In2Se3 Nanoflakes},
journal = {Nano Letters} ,
volume = {17},
pages = {5508},
year = {2017}
}

@Article{Wan2018-exp,
author ="Wan, Siyuan and Li, Yue and Li, Wei and Mao, Xiaoyu and Zhu, Wenguang and Zeng, Hualing",
title  ="Room-temperature ferroelectricity and a switchable diode effect in two-dimensional α-In2Se3 thin layers",
journal  ="Nanoscale",
year  ="2018",
volume  ="10",
issue  ="31",
pages  ="14885-14892",
publisher  ="The Royal Society of Chemistry"}

@article{Xue2018-exp,
author = {Xue, Fei and Hu, Weijin and Lee, Ko-Chun and Lu, Li-Syuan and Zhang, Junwei and Tang, Hao-Ling and Han, Ali and Hsu, Wei-Ting and Tu, Shaobo and Chang, Wen-Hao and Lien, Chen-Hsin and He, Jr-Hau and Zhang, Zhidong and Li, Lain-Jong and Zhang, Xixiang},
title = {Room-Temperature Ferroelectricity in Hexagonally Layered α-In2Se3 Nanoflakes down to the Monolayer Limit},
year = {2018},
journal = {Advanced Functional Materials},
volume = {28},
number = {50},
pages = {1803738}
}

@article{Cui2018-exp-gap,
author = {Chaojie Cui and  Wei-Jin Hu and Xingxu Yan and Christopher Addiego and Wenpei Gao and Yao Wang and Zhe Wang and Linze Li and Yingchun Cheng and Peng Li and Xixiang Zhang and Husam N. Alshareef and Tom Wu and Wenguang Zhu and Xiaoqing Pan and Lain-Jong Li},
title = {Intercorrelated In-Plane and Out-of-Plane Ferroelectricity in Ultrathin Two-Dimensional Layered Semiconductor In2Se3},
year = {2018},
journal = {Nano Letters},
volume = {18},
pages = {1253}
}

@article{Xiao2018,
title = {Intrinsic Two-Dimensional Ferroelectricity with Dipole Locking},
author = {Xiao, Jun and Zhu, Hanyu and Wang, Ying and Feng, Wei and Hu, Yunxia and Dasgupta, Arvind and Han, Yimo and Wang, Yuan and Muller, David A. and Martin, Lane W. and Hu, PingAn and Zhang, Xiang},
journal = {Phys. Rev. Lett.},
volume = {120},
issue = {22},
pages = {227601},
numpages = {6},
year = {2018}
}

@article{Wang2018,
  title = {Layer and doping tunable ferromagnetic order in two-dimensional $\mathrm{Cr}{\mathrm{S}}_{2}$ layers},
  author = {Wang, Cong and Zhou, Xieyu and Pan, Yuhao and Qiao, Jingsi and Kong, Xianghua and Kaun, Chao-Cheng and Ji, Wei},
  journal = {Phys. Rev. B},
  volume = {97},
  issue = {24},
  pages = {245409},
  numpages = {11},
  year = {2018},
  month = {Jun},
  publisher = {American Physical Society}
}

@article{Armitage2025,
  title = {Electronic structure of monolayer ${\mathrm{CrTe}}_{2}$: An antiferromagnetic two-dimensional van der Waals material},
  author = {Armitage, Olivia and Kushwaha, Naina and Rajan, Akhil and Rhodes, Luke C. and Buchberger, Sebastian and Saika, Bruno Kenichi and Mo, Shu and Watson, Matthew D. and King, Phil D. C. and Wahl, Peter},
  journal = {Phys. Rev. B},
  volume = {112},
  issue = {24},
  pages = {245416},
  numpages = {9},
  year = {2025},
  month = {Dec},
  publisher = {American Physical Society}
}

@article{kuppers2018,
    author = {K{\"u}pers, Michael and Konze, Philipp M. and Meledin, Alexander and Mayer, Joachim and Englert, Ulli and Wuttig, Matthias and Dronskowski, Richard},
    title = {Controlled Crystal Growth of Indium Selenide, In2Se3, and the Crystal Structures of α-In2Se3},
    journal = {Inorganic Chemistry},
    volume = {57},
    number = {18},
    pages = {11775-11781},
    year = {2018},
    month = {09},
    issn = {0020-1669}
}

@article{Balakrishnan_2018,
    year = {2018},
    month = {jun},
    publisher = {IOP Publishing},
    volume = {5},
    number = {3},
    pages = {035026},
    author = {Balakrishnan, Nilanthy and Steer, Elisabeth D and Smith, Emily F and Kudrynskyi, Zakhar R and Kovalyuk, Zakhar D and Eaves, Laurence and Patanè, Amalia and Beton, Peter H},
    title = {Epitaxial growth of γ-InSe and α, β, and γ-In2Se3 on ε-GaSe},
    journal = {2D Materials}
}

@Article{Wei2019,
author ="Sun, Wei and Wang, Wenxuan and Chen, Dong and Cheng, Zhenxiang and Wang, Yuanxu",
title  ="Valence mediated tunable magnetism and electronic properties by ferroelectric polarization switching in 2D FeI2/In2Se3 van der Waals heterostructures",
journal  ="Nanoscale",
year  ="2019",
volume  ="11",
issue  ="20",
pages  ="9931-9936"
}

@article{Cheng2019,
    author = "Gong, Cheng and Kim, Eun Mi and Wang, Yuan and Lee, Geunsik and Zhang, Xiang",
    title = "Multiferroicity in atomic van der Waals heterostructures",
    journal = "Nature Communications" ,
    volume = "10",
    pages = "2657",
    year = {2019},
}

@Article{Ting2020,
author ="Hu, Ting and Su, Haishan and Li, Qiongyu and Kan, Erjun",
title  ="Tunable ferroelectric single-atom catalysis of CO oxidation using a Pt/In2Se3 monolayer",
journal  ="J. Mater. Chem. A",
year  ="2020",
volume  ="8",
issue  ="39",
pages  ="20725-20731",
}

@article{Picozzi2020,
author = {Shang, Jing and Tang, Xiao and Gu, Yuantong and Krasheninnikov, Arkady V. and Picozzi, Silvia and Chen, Changfeng and Kou, Liangzhi},
title = {Robust Magnetoelectric Effect in the Decorated Graphene/In2Se3 Heterostructure},
journal = {ACS Applied Materials \& Interfaces},
volume = {13},
number = {2},
pages = {3033-3039},
year = {2021},
}

@article{Wang2021,
    author = {Wang, Wenxuan and Sun, Wei and Li, Hang and Bai, Ying and Ren, Fengzhu and You, Caiyin and Cheng, Zhenxiang},
    title = {Nonvolatile magnetoelectric coupling in two-dimensional ferromagnetic-bilayer/ferroelectric van der Waals heterostructures},
    journal = {Nanoscale},
    volume = {13},
    number = {33},
    pages = {14214-14220},
    year = {2021},
    month = {09}
}

@article{Ju2021,
author = {Ju, Lin and Tan, Xin and Mao, Xin and Gu, Yuantong and Smith, Sean and Du, Aijun and Chen, Zhongfang and Chen, Changfeng and Kou, Liangzhi},
title = {Controllable CO2 electrocatalytic reduction via ferroelectric switching on single atom anchored In2Se3 monolayer},
journal = {Nature Communications},
volume = {12},
pages = {5128},
year = {2021},
}

@article{Zang2024,
title = {Electronic structure, Curie temperature, and magnetic transport of a two-dimensional multiferroic ${\text{MnSe}}_{2}/{\text{In}}_{2}{\text{Se}}_{3}$ heterostructure},
author = {Zhang, Zi-Wen and Lang, Yu-Fei and Zhu, Hui-Ping and Li, Bo and Zhao, Yu-Qing and Wei, Bin and Zhou, Wu-Xing},
journal = {Phys. Rev. Appl.},
volume = {21},
issue = {6},
pages = {064012},
numpages = {11},
year = {2024},
}

@article{slater1954,
author = {J. C. Slater and G. F. Koster},
journal = {Phys. Rev.},
volume = {94},
pages = {1498},
year = {1954},
}

@article{souza2001,
author = {I. Souza, N. Marzari and D. Vanderbilt},
journal = {Phys. Rev. B},
volume = {65},
pages = {035109},
year = {2001},
}

@article{wannier90_2014,
  author   = {Mostofi, Arash A. and Yates, Jonathan R. and Pizzi, Giovanni and Lee, Young-Su and Souza, Ivo and Vanderbilt, David and Marzari, Nicola},
  title    = {An updated version of {Wannier90}: {A} tool for obtaining maximally-localised {Wannier} functions},
  journal  = {Computer Physics Communications},
  volume   = {185},
  number   = {8},
  pages    = {2309--2310},
  year     = {2014},
  doi      = {10.1016/j.cpc.2014.05.003}
}

@article{wannier90_2020,
  author   = {Pizzi, Giovanni and Vitale, Valerio and Arita, Ryotaro and Bl{\"u}gel, Stefan and Freimuth, Frank and G{\'e}ranton, Guillaume and Gibertini, Marco and Gresch, Dominik and Johnson, Charles and Koretsune, Takashi and Iba{\~n}ez-Azpiroz, Julen and Lee, Hyungjun and Lihm, Jae-Mo and Marchand, Daniel and Marrazzo, Antimo and Mokrousov, Yuriy and Mustafa, Jamal I. and Nohara, Yoshiro and Nomura, Yusuke and Paulatto, Lorenzo and Ponc{\'e}, Samuel and Ponweiser, Thomas and Qiao, Junfeng and Th{\"o}le, Florian and Tsirkin, Stepan S. and Wierzbowska, Ma{\l}gorzata and Marzari, Nicola colored and Vanderbilt, David and Souza, Ivo and Mostofi, Arash A. and Yates, Jonathan R.},
  title    = {{Wannier90} as a community code: new features and applications},
  journal  = {Journal of Physics: Condensed Matter},
  volume   = {32},
  number   = {40},
  pages    = {405902},
  year     = {2020},
  doi      = {10.1088/1361-648X/ab92a9}
}

@article{Hubbard2022,
  title = {Hubbard $U$ parameters for transition metals from first principles},
  author = {Tesch, Rebekka and Kowalski, Piotr M.},
  journal = {Phys. Rev. B},
  volume = {105},
  issue = {19},
  pages = {195153},
  numpages = {13},
  year = {2022},
  month = {May},
  publisher = {American Physical Society},
  doi = {10.1103/PhysRevB.105.195153},
  url = {https://link.aps.org/doi/10.1103/PhysRevB.105.195153}
}

@article{Hubbard2025,
title = {Hubbard U parameters for transition metals: Correction of the Bloch state and atomic values},
journal = {Computational Materials Science},
volume = {258},
pages = {114064},
year = {2025},
issn = {0927-0256},
doi = {https://doi.org/10.1016/j.commatsci.2025.114064},
url = {https://www.sciencedirect.com/science/article/pii/S0927025625004070},
author = {Jacques R. Eone}
}

@book{Harrison1980,
  author    = {Harrison, Walter A.},
  title     = {Electronic Structure and the Properties of Solids:
               The Physics of the Chemical Bond},
  publisher = {W. H. Freeman},
  address   = {San Francisco},
  year      = {1980}
}

\clearpage
\includepdf[
    pages=-,
    pagecommand={\thispagestyle{empty}}
]{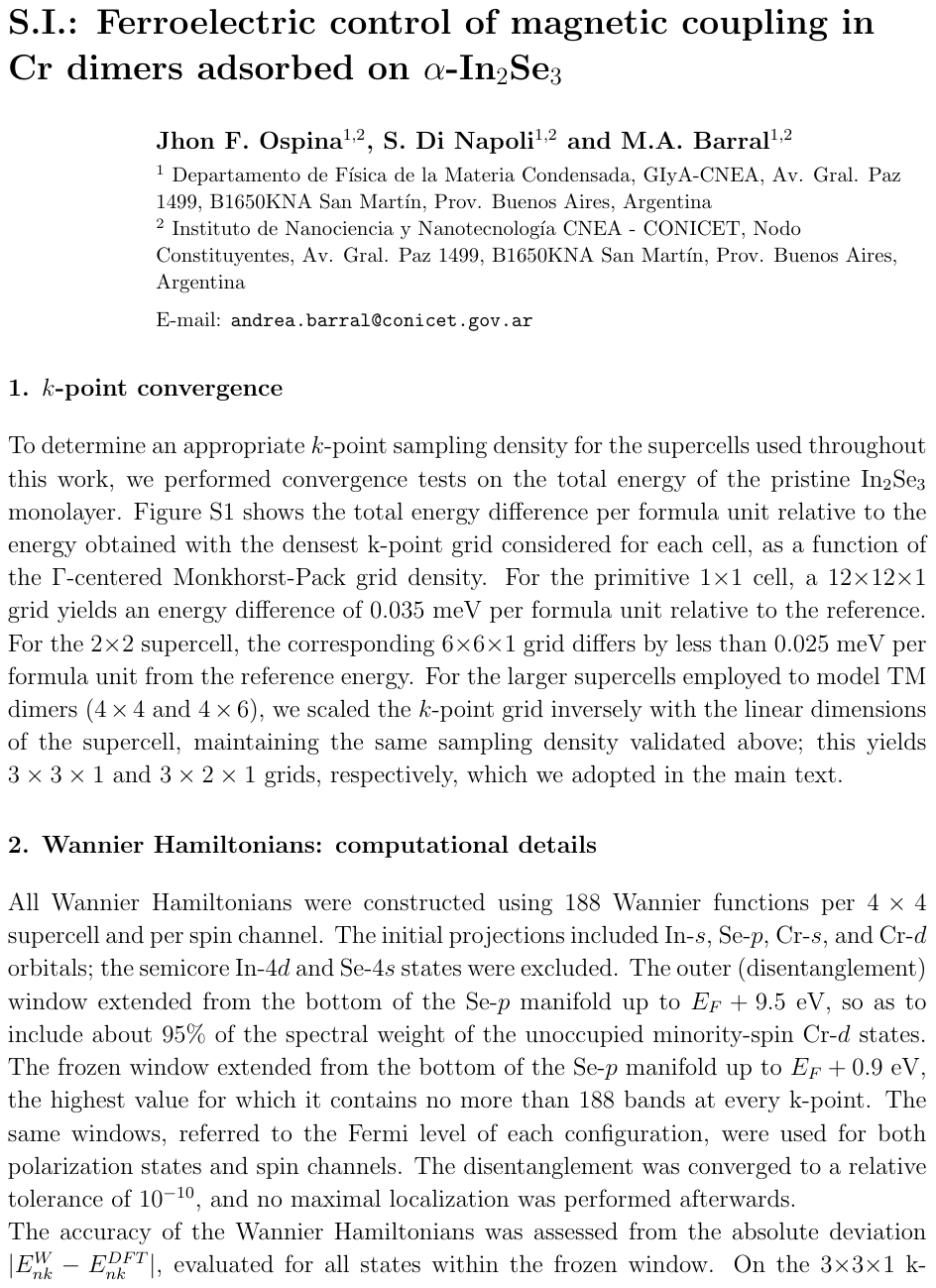}

\end{document}